\documentclass[aps,prl,preprint,amsmath,amssymb,showpacs,superscriptaddress]{revtex4-1}
\usepackage{graphicx}
\usepackage{dcolumn}
\usepackage{bm}
\usepackage{epstopdf}

\usepackage{hyperref}

\begin{document}


\title{Symmetry energy at supra-saturation densities via the Gravitational Waves from GW170817}

\author{Hui Tong 
}
\affiliation{State Key Laboratory of Nuclear Physics and Technology, School of Physics, Peking University \\
Beijing 100871, China}

\author{Peng-Wei Zhao 
}
\affiliation{State Key Laboratory of Nuclear Physics and Technology, School of Physics, Peking University \\
Beijing 100871, China}

\author{Jie Meng 
}
\affiliation{State Key Laboratory of Nuclear Physics and Technology, School of Physics, Peking University \\
Beijing 100871, China}
\affiliation{Department of Physics, University of Stellenbosch, Stellenbosch, South Africa}
\affiliation{Yukawa Institute for Theoretical Physics, Kyoto University, Kyoto 606-8502, Japan}


\begin{abstract}
Motivated by the historical detection of gravitational waves from GW170817, the neutron star and the neutron drop, i.e., a certain number of neutrons confined in an external field, are systematically investigated by \emph{ab initio} calculations as well as the nonrelativistic and relativistic state-of-art density functional theories.
Strong correlations are found among the neutron star tidal deformability, the neutron star radius, the root-mean-square radii of neutron drops, and the symmetry energies of nuclear matter at supra-saturation densities.
From these correlations and the upper limit on the tidal deformability extracted from GW170817, the neutron star radii, the neutron drop radii, and the symmetry energy at twice saturation density are respectively constrained as $R_{1.4M_{\odot}}\leqslant 12.94$~km, $R_{\rm nd} \leqslant 2.36$~fm, and $E_{\mathrm{sym}}(2\rho_0) \leqslant 53.2$~MeV.
\end{abstract}



\maketitle




On August 17, 2017, a gravitational wave (GW) signal from a merger of binary neutron star (BNS) system was observed by the advanced LIGO and Virgo Collaborations for the first time, i.e., GW170817~\cite{Abbott2017}.
This observation, together with the followed observation of the associated electromagnetic emissions~\cite{VSavchenko2017,A.Goldstein2017}, has opened a significant multimessenger era in the field of astrophysics and nuclear physics.
In particular, the GW signal provides a crucial opportunity to investigate the properties of the compact matter---neutron star.

According to the GW170817 signal from the merger of BNS, one can extract the masses of neutron stars.
The measured chirp mass~\cite{Peters1963} is $\mathcal{M}=1.188_{-0.002}^{+0.004}M_{\odot}$~\cite{Abbott2017}, which is defined as $\mathcal{M}=(M_1M_2)^{3/5}(M_1+M_2)^{-1/5}$, from the component masses $M_1$ and $M_2$.
In addition, the tidal deformability $\Lambda$ can also be extracted, which represents the mass quadrupole moment response of a neutron star to the strong gravitational field induced by its companion.
The tidal deformability of a neutron star is defined as  $\Lambda=\frac{2}{3}k_2\left(\frac{c^2R}{GM}\right)^5$, where $k_2$~is the second Love number, $M$ and $R$ are the neutron star mass and radius, $c$ and $G$ are the speed of light and the gravitational constant, respectively~\cite{Damour1992,Hinderer2008,Flanagan2008,Damour2009,Postnikov2010}.
Therefore, the tidal deformability is expected to have a high sensitivity to the neutron star radius ($\Lambda \sim R^5$).
For a neutron star with mass $1.4M_{\odot}$ , the LIGO and Virgo collaborations provided an upper limit $\Lambda_{1.4M_{\odot}} \leqslant 580$ with a credible level of 90\%~\cite{Abbott2018}.

Both the mass and tidal deformability of neutron stars provide new opportunities to study the equation of state (EOS) for nuclear matter~\cite{Lattimer2004,Lattimer2016}. In particular, its isovector properties, e.g., symmetry energy, play a significant role in exploring new physics in both nuclear physics and astrophysics.
By adopting the tidal deformability limit from the GW signal, the neuron star radius as well as the neutron skin thickness of $^{208}$Pb, which is strongly correlated to the slope of the symmetry energy $L$ at nuclear matter saturation density ($\rho_0\approx0.16$~fm$^{-3}$), have been inferred~\cite{Fattoyev2018}.
A direct relation between the tidal deformability and $L$ has been studied in Refs.~\cite{Lim2018,Malik2018}.
However, both works indicate that the tidal deformability, which probes the symmetry energy at around twice saturation density, may not provide a strong constraint on the nuclear matter bulk parameters at the saturation density.

A neutron drop system is very useful to probe the high-density behavior of nuclear matter.
The neutron drop is a novel inhomogeneous pure neutron system consisting of a few neutrons confined in an external field.
By varying the number of neutrons and/or the strength of the external field for neutron drops, one can obtain
a variety of information for the drops at various densities. The nuclear EOS ranging from lower densities to higher densities could be correlated to the neutron drop properties~\cite{Zhao2016}.

An accurate knowledge of the neutron drop properties is crucial to understand the physics of neutron stars~\cite{Ravenhall1983,Brown2009,Buraczynski2016} and to probe possible new physics for nuclei with large isospin in as yet unexplored regions of the nuclear chart~\cite{Erler2012}.
Due to its simplicity, the neutron drop is widely used to test various nuclear many-body techniques, e.g., \emph{ab initio} approaches for light nuclei and density functional theories (DFTs) for heavy ones~\cite{Pudliner1996,Gandolfi2011,Bogner2011,Maris2013,Potter2014,Carlson2015,Zhao2016,shen2018,shen20182}.
The former are based on realistic nucleon-nucleon ($NN$) and three-nucleon ($3N$) interactions with a many-body Hamiltonian, while the latter depend on a variation of an energy functional with respect to nucleon densities.

Therefore, it is timely and interesting to link the GW signals to the nuclear EOS and the neutron drop properties, so as to promote both the nuclear physics and the astrophysics with the advent of the multi-messenger era.

In this Letter, the neutron star tidal deformability and its correlation with the neutron star radius are systematically investigated with the state-of-art DFTs~\cite{RING1996,Bender2003,Vretenar2005,CDFT2016} and the \emph{ab initio} relativistic Brueckner-Hartree-Fock (RBHF) theory~\cite{Krastev2006,shen2016,Tong2018}.
A strong linear correlation between the root-mean-square (rms) radii of neutron drops and the neutron star radii is demonstrated.
Furthermore, a strong correlation between the symmetry energy at twice saturation density and the neutron drop radii is illustrated.
With these correlations, the upper limit of the tidal deformability $\Lambda_{1.4M_{\odot}} \leqslant 580$~\cite{Abbott2018} extracted from GW170817 is used to constrain the rms radii of neutron drops and, in turn, the symmetry energies of nuclear matter at supra-saturation densities.

\begin{figure}[!htbp]
  \centering
  \includegraphics[width=8.7cm]{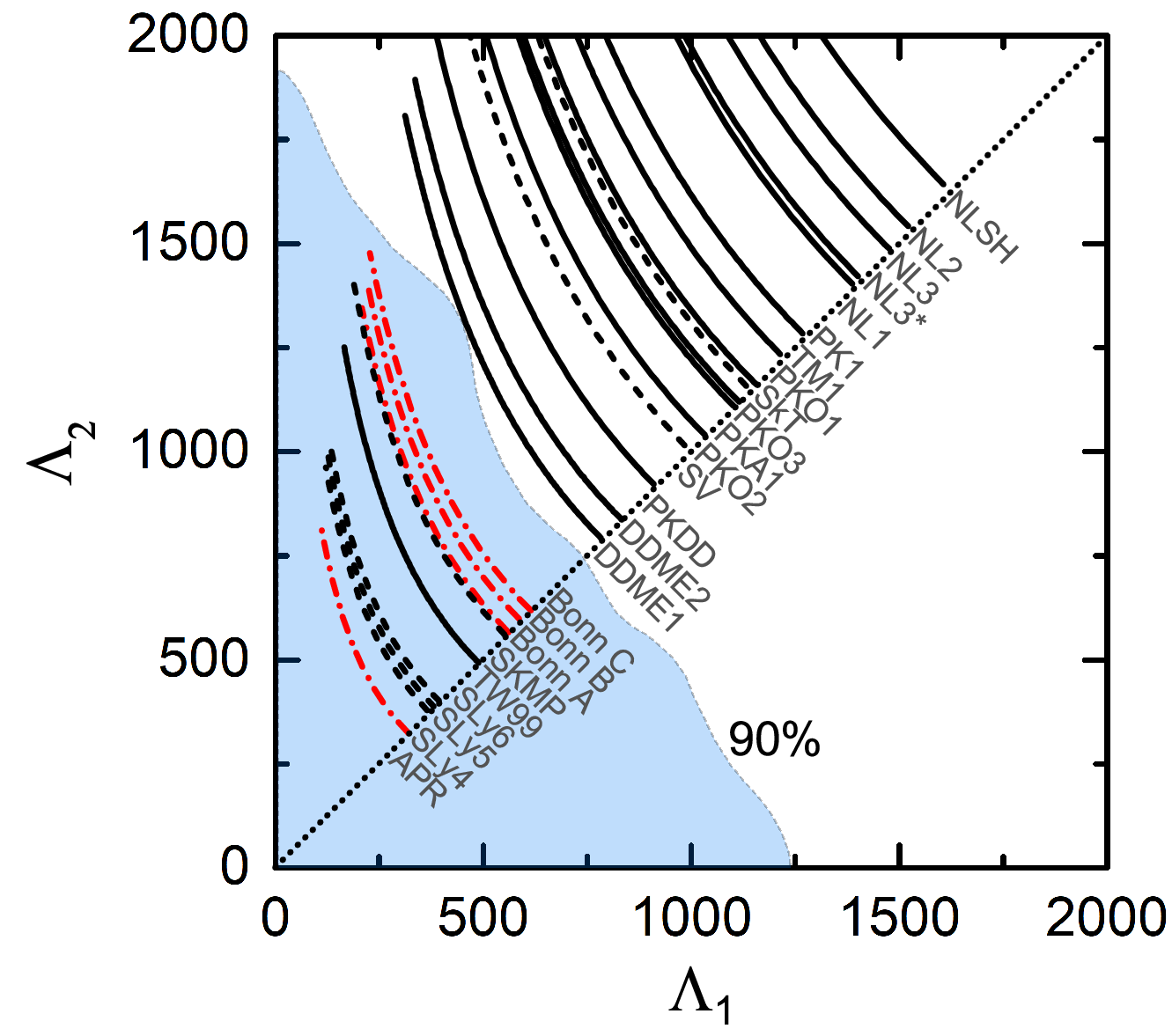}
  \caption{(Color online.)
  Tidal deformabilities $\Lambda_1$ and $\Lambda_2$ associated with the high mass $M_1$ and low mass $M_2$ components of the binary neutron star obtained by a set of relativistic (solid lines), nonrelativistic (dashed lines) density functionals, and the results from \emph{ab initio} Relativistic Brueckner-Hartree-Fock (RBHF) theory using the Bonn potentials and the \emph{ab initio} variational calculations (APR)~\cite{Akmal1998} (dot-dashed lines). The
  boundary of $\Lambda_1=\Lambda_2$ is denoted by the diagonal dotted
  line. The 90\% probability contour is extracted from GW170817~\cite{Abbott2018}.
  }
  \label{Fig1}
\end{figure}

In the context of GW170817, the tidal deformabilities $\Lambda_1$ and $\Lambda_2$ associated with the high mass $M_1$ and low mass $M_2$ components of the BNS are calculated by a series of DFTs and the \emph{ab initio} RBHF theory.
As shown in Fig.~\ref{Fig1}, the calculations here are carried out by many well-determined density functionals widely used for nuclear and astrophysical physics. In particular, the EOSs here are consistent with the observed lower bound on the maximum mass of neutron stars for $M_{\text{max}}>2M_{\odot}$~\cite{Demorest2010,Antoniadis2013}.
The adopted DFTs range from nonrelativistic models (e.g., SLy4 and those starting with S)~\cite{Dutra2012}, to relativistic models with the nonlinear meson exchange functionals (NL models, PK1, TM1), as well as the density-dependent meson exchange ones (DD-ME and RHF models, PKDD, TW99)~\cite{Dutra2014,Long2006}.
\emph{Ab initio} calculations have been carried out with the RBHF theory using the Bonn potentials~\cite{Tong2018}.
Note that the \emph{ab initio} variational calculations (APR) are also shown for comparison~\cite{Akmal1998}.

Allowing a variation of the high mass $M_1$ within the range $1.365M_{\odot} \leqslant M_1 \leqslant 1.60M_{\odot}$, the corresponding low mass $M_2$ can be  determined by the chirp mass $\mathcal{M}=(M_1M_2)^{3/5}(M_1+M_2)^{-1/5}=1.188M_{\odot}$ given by GW170817~\cite{Abbott2017}, and one can then obtain the relations between $\Lambda_1$ and $\Lambda_2$ of a BNS.
As a comparison, we also plot the latest $90\%$ probability contour recommended by the LIGO and Virgo observatories with the low-spin priors~\cite{Abbott2018}.
Since the isovector channels of the density functionals shown in Fig.~\ref{Fig1} are loosely determined in the fitting procedures, their predictions span a fairly wide range of tidal deformabilities $\Lambda$.
For a given neutron star mass, the functionals with a softer (stiffer) symmetry energy yield smaller (larger) stellar radii and correspondingly smaller (larger) tidal deformabilities.
One can see that the curves locating on the right side of the contour are ruled out by GW170817.
This indicates that overly stiff EOSs, which produce large values of $\Lambda$ and correspondingly large radii for fixed $M$ are disfavored by the GW170817.
Nevertheless, the \emph{ab initio} calculations and a relatively large set of EOSs situate inside the $90\%$ probability contour.

\begin{figure}[htbp]
  \centerline{
  \includegraphics[width=8.8cm]{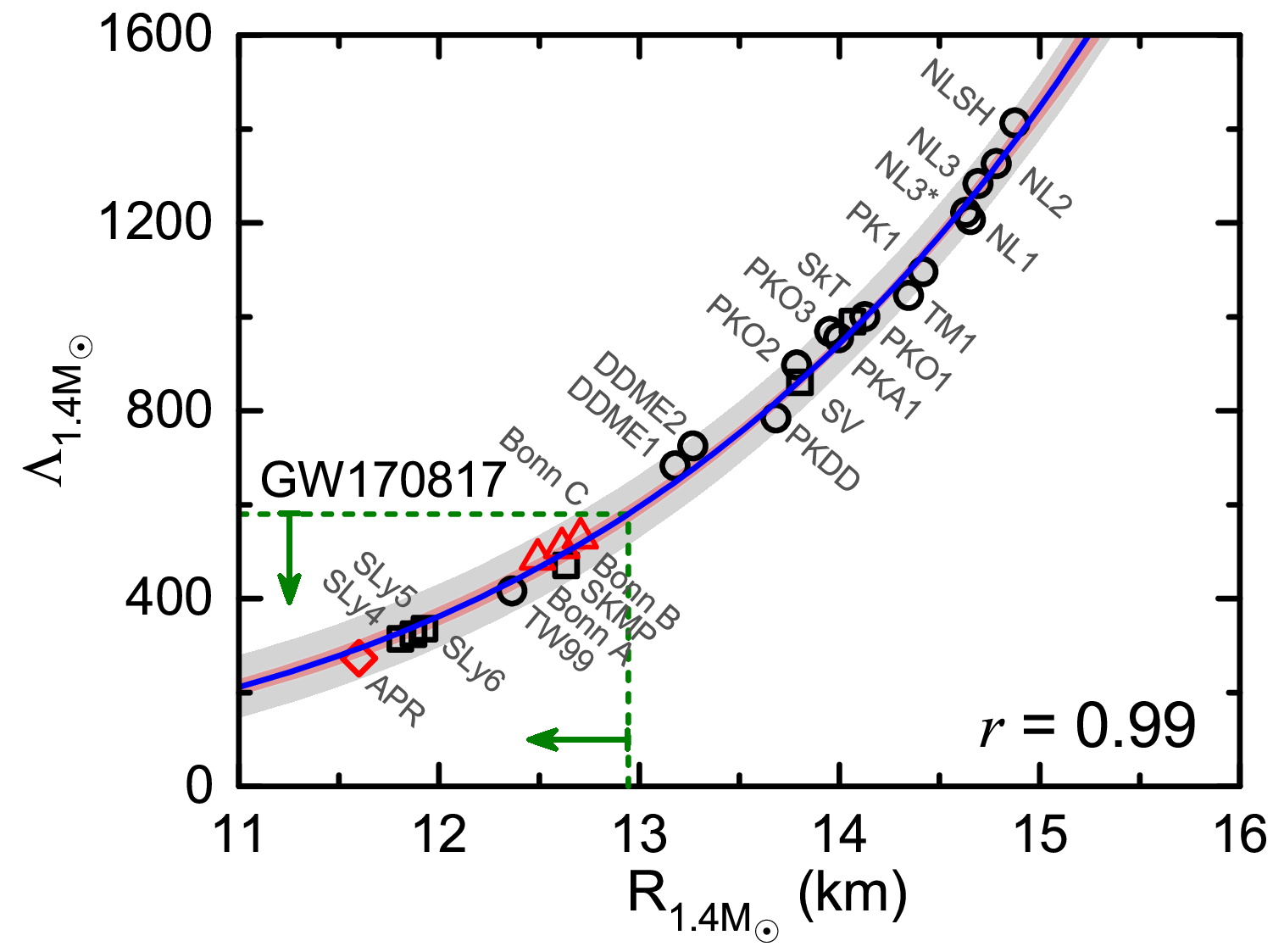}
  }
  \caption{(Color online)
  The tidal deformability $\Lambda_{1.4M_{\odot}}$ of a 1.4$M_{\odot}$ neutron star as a function of the corresponding radius $R_{1.4M_{\odot}}$(km) as predicted by various relativistic (circles) and nonrelativistic (squares) density functionals, in comparison with results calculated by RBHF theory using the Bonn potentials (triangles) and the \emph{ab initio} variational calculations (APR, diamond).
  The blue curve line is the fit to the results of density functionals, and the inner (outer) colored regions depict the 95\% confidence (prediction) intervals of the curvilinear regression.
  The vertical arrow indicates the constraint $\Lambda_{1.4M_{\odot}} \leqslant 580$~\cite{Abbott2018} from GW170817, which leads to $R_{1.4M_{\odot}} \leqslant 12.94$~km labeled by the horizontal arrow according to the correlations.
  }
  \label{Fig2}
\end{figure}

The tidal deformabilities predicted by both relativistic and nonrelativistic DFTs are very useful to extract the correlations between the tidal deformabilities and other isovector-sensitive properties.
The correlations between the tidal deformability $\Lambda_{1.4M_{\odot}}$ and the radius $R_{1.4M_{\odot}}$ of a neutron star with mass $M =1.4M_{\odot} $
are shown in Fig.~\ref{Fig2}.
The predictions of DFTs are fitted by the equation $\Lambda_{1.4M_{\odot}}=aR_{1.4M_{\odot}}^b$ with $a\approx7.29\times10^{-5}$(km$^{-b}$) and $b\approx6.21$. A strong correlation between $\Lambda_{1.4M_{\odot}}$ and $R_{1.4M_{\odot}}$ with a correlation coefficient of $r\approx0.99$ is revealed.

It is important to underscore that this strong correlation between $\Lambda_{1.4M_{\odot}}$ and $R_{1.4M_{\odot}}$ is universal since it is based
on widely different nuclear density functionals.
Furthermore, the correlation predicted by DFTs is supported by the \emph{ab initio} calculations including the RBHF theory and the APR, as shown in Fig.~\ref{Fig2}.
The universal correlation between $\Lambda_{1.4M_{\odot}}$ and $R_{1.4M_{\odot}}$ allows one to extract $R_{1.4M_{\odot}}$ from the tidal deformability.
From GW170817, the result $\Lambda_{1.4M_{\odot}}\leqslant 580$~\cite{Abbott2018} leads to $R_{1.4M_{\odot}}\leqslant 12.94$~km.
This is consistent with other works, such as $R_{1.4M_{\odot}}\leqslant 13.76$~km~\cite{Fattoyev2018} and $R_{1.4M_{\odot}}\leqslant 13.6$~km~\cite{Annala2018} from the tidal deformability~\cite{Abbott2017}, and $R_{1.4M_{\odot}}\leqslant 13.6$~km from the available terrestrial laboratory data~\cite{LI2006}.

The neutron drop is an inhomogeneous pure neutron system consisting of a few neutrons confined in an external field.
Although the radius of a neutron drop is smaller than that of a neutron star by 19 orders of magnitude, both quantities highly depend on the symmetry energy~\cite{LATTIMER2000,Gandolfi2012,BALDO2016,Zhao2016}.
Therefore, it is interesting to investigate the correlations between them.

\begin{figure}[htbp]
  \centerline{
  \includegraphics[width=9.0cm]{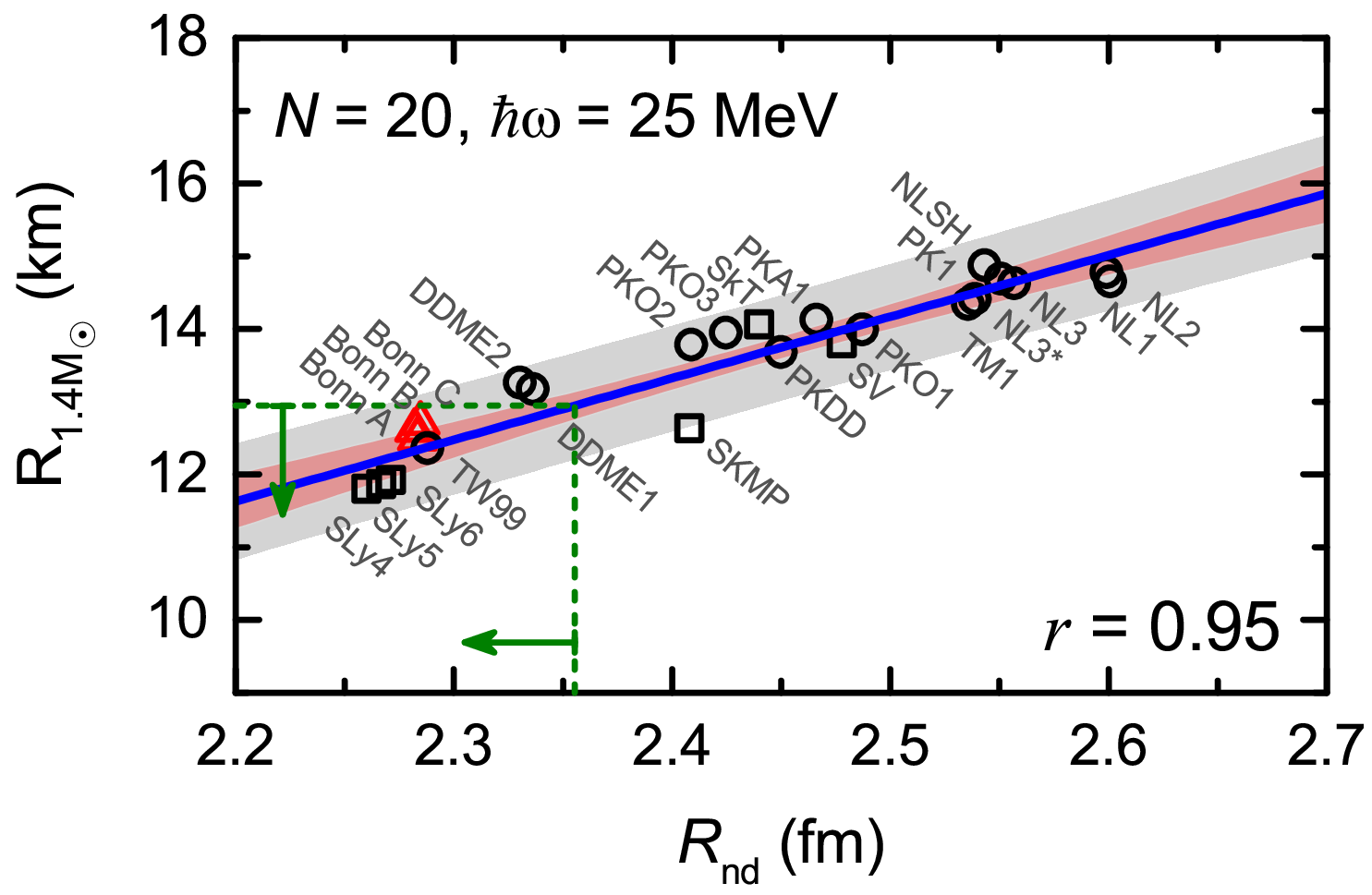}
  }
  \caption{(Color online)
  The radius of a 1.4$M_{\odot}$ neutron star against the root-mean-square (rms) radius $R_{\rm nd}$
  for $N = 20$ neutrons trapped in a HO potential ($\hbar \omega$ = 25 MeV) obtained with various relativistic (circles) and nonrelativistic (squares) density functionals, as well as the RBHF theory using the Bonn potentials (triangles).
  The blue line is a linear fit to the results of density functionals.
  The upper limit $R_{1.4M_{\odot}} \leqslant 12.94$~km deduced from Fig.~\ref{Fig2} is shown with the vertical arrow, which gives a constraint $R_{\rm nd} \leqslant 2.36$~fm labeled by the horizontal arrow.
  }
  \label{Fig3}
\end{figure}

In Fig.~\ref{Fig3}, the radius of a 1.4$M_{\odot}$ neutron star and the rms radius $R_{\rm nd}$ for $N = 20$ neutrons confined in a harmonic oscillator (HO) potential ($\hbar \omega$ = 25 MeV) calculated with relativistic and nonrelativistic DFTs and the \emph{ab initio} RBHF theory are given.
The DFTs predictions for $R_{1.4M_{\odot}}$ and $R_{\rm nd}$ respectively range from 11.5 km to 15.0 km, and from 2.25 fm to 2.60 fm, due to their differences in the isovector channels.

A strong linear correlation can be established between the neutron star radius $R_{1.4M_{\odot}}$ and the rms radius of the neutron drop $R_{\rm nd}$ by fitting the results from DFTs. The Pearson's correlation coefficient is $r = 0.95$.
This linear correlation between $R_{1.4M_{\odot}}$ and $R_{\rm nd}$ is universal with the neutron number and the strength of the external field for the neutron drops, as long as the central density of the drop is comparable with that of the neutron stars, i.e., around 2--3 times saturation density.

The correlation between $R_{1.4M_{\odot}}$ and $R_{\rm nd}$ predicted by DFTs is supported by the \emph{ab initio} RBHF calculations.
The verification of this linear correlation from \emph{ab initio} calculations gives us confidence to deduce $R_{\rm nd} \leqslant 2.36$~fm from $R_{1.4M_{\odot}} \leqslant 12.94$~km obtained with the $\Lambda_{1.4M_{\odot}}$ vs $R_{1.4M_{\odot}}$ relation in Fig.~\ref{Fig2}.
With the future high-accuracy GW measurements, the constraints on the radii of neutron drops would be accurately deduced, which might provide a useful constraint for three-neutron forces~\cite{Zhao2016}.

Apart from the constraint from GW170817, the correlation can also be used in combination with other astronomical observations. For example, the future precise radii of neutron stars from photospheric radius expansion X-ray bursts and quiescent low-mass X-ray binaries~\cite{Steiner2010,Bogdanov2016} can provide a similar constraint on the neutron drop radii.

By changing the strength of the external field, the neutron drop system with a central density up to 3 times saturation density can be obtained, which is very useful to probe the symmetry energy at high density.

The symmetry energy at the density $\rho$ is defined as $E_{\mathrm{sym}}(\rho)=\frac{1}{2}\frac{\partial^{2}E(\rho,\alpha)}
{\partial\alpha^{2}}\big|_{\alpha=0}$, where $E(\rho,\alpha)$ is the binding energy per nucleon and the asymmetry parameter $\alpha = \frac{\rho_n-\rho_p}{\rho}$ with the proton densities $\rho_p$ and the neutron densities $\rho_n$.
The density dependence of the nuclear symmetry energy is essential to understand both finite nuclei and neutron stars.
However, the constraints on the density dependence of the symmetry energy either from the experiment of the heavy ion collisions~\cite{Tsang2009} or the predictions from variant models~\cite{Li2008} have large ambiguity.

\begin{figure}[htbp]
  \centerline{
  \includegraphics[width=9.0cm]{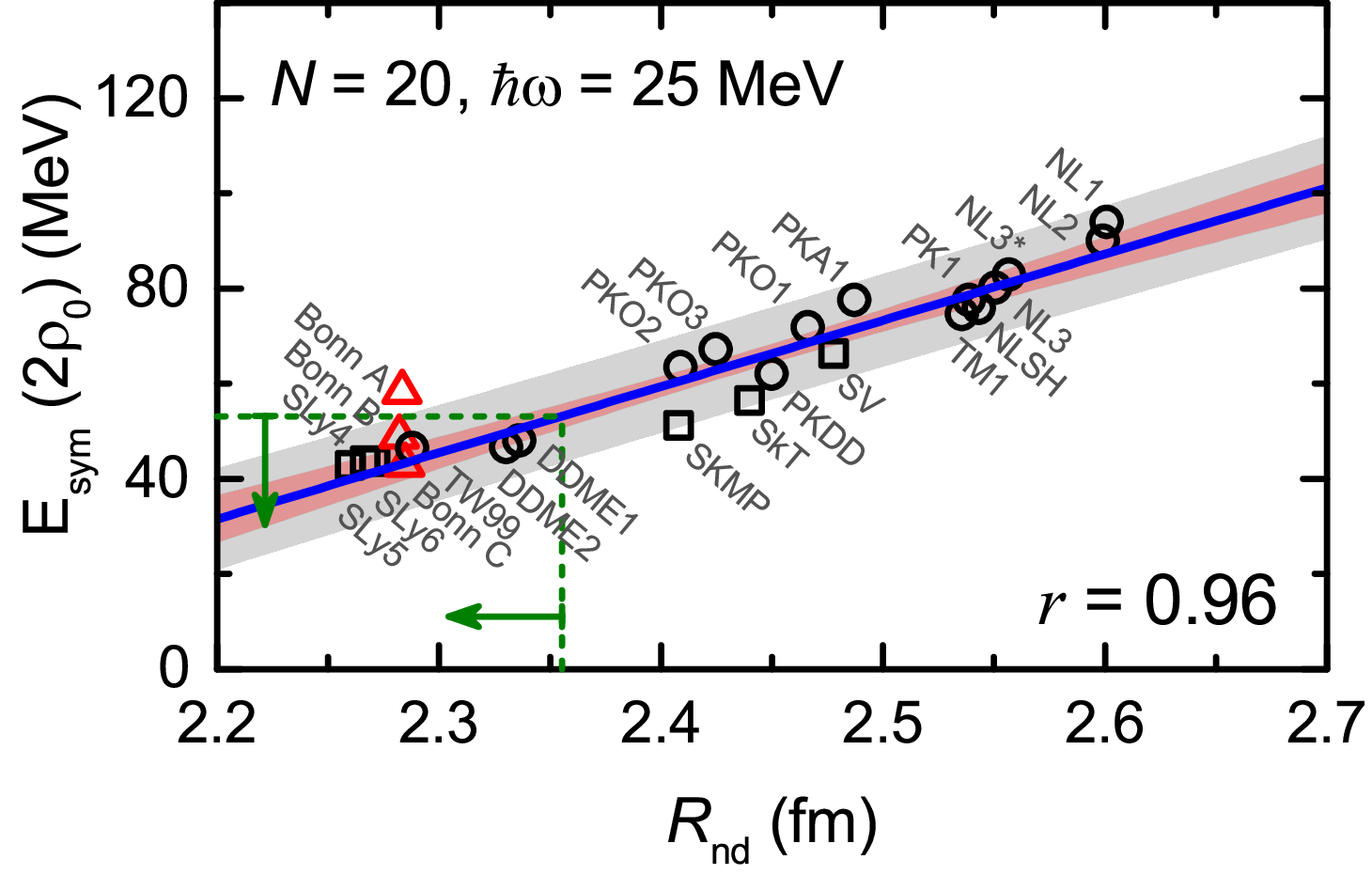}
  }
  \caption{(Color online)
  Correlations between the symmetry energy of nuclear matter at twice nuclear saturation density and rms radii of 20 neutrons in a HO with $\hbar\omega = 25$~MeV, calculated by relativistic (circles), nonrelativistic (squares) density functional theories and \emph{ab initio} methods (triangles).
  The upper limit $R_{\rm nd} \leqslant 2.36$~fm deduced from the correlations between $R_{1.4}$ vs $R_{\rm nd}$ is shown with the horizontal arrow, which gives rise to an upper limit $E_{\mathrm{sym}}(2\rho_0) \leqslant 53.2$~MeV labeled by the vertical arrow.
  }
  \label{Fig4}
\end{figure}

In Fig.~\ref{Fig4}, the symmetry energy of nuclear matter at twice saturation density $E_{\mathrm{sym}}(2\rho_0)$ and the rms radii $R_{\rm nd}$ of 20 neutrons in a HO with $\hbar\omega = 25$~MeV calculated by DFTs are given.
A strong correlation can be established between $E_{\mathrm{sym}}(2\rho_0)$ and the neutron drop radius $R_{\rm nd}$.
The Pearson's coefficient by fitting the DFTs results is $r = 0.96$.
It demonstrates that a precise determination of the neutron drop radii could set a basic constraint on the symmetry energy at twice saturation density.
The constraint $R_{\rm nd} \leqslant 2.36$~fm in Fig.~\ref{Fig3} yields $E_{\mathrm{sym}}(2\rho_0) \leqslant 53.2$~MeV.

The constraint on the symmetry energy $E_{\mathrm{sym}}(2\rho_0) \leqslant 53.2$~MeV is consistent with the recent ASY-EOS experimental results $50.8$~MeV $\leqslant E_{\mathrm{sym}}(2\rho_0) \leqslant$ $60.4$~MeV~\cite{Russotto2016}
and the predictions by \emph{ab initio} RBHF theory.
Accordingly, the ambiguity of the symmetry energy predicted at twice saturation density is remarkably reduced.
The present results provide an excellent agreement between the astrophysical GW observations and the terrestrial
experiments, and reveal a soft symmetry energy in the regime of high density.

Similar correlations between $R_{\mathrm{nd}}$ and $E_{\mathrm{sym}}(\rho)$ for $\rho=\rho_0\sim4\rho_0$ have been analyzed.
The symmetry energies from $E_{\mathrm{sym}}(2\rho_0)$ to $E_{\mathrm{sym}}(3\rho_0)$ are correlated with $R_{\mathrm{nd}}$ in a linear way with the Pearson's coefficient $r\sim0.96$.
Such a connection should be reasonable because the central density of the neutron drop with 20 neutrons confined in a HO potential with $\hbar \omega = 25$ MeV is around and even slightly larger than the twice saturation density.
Although there is no data for the symmetry energy beyond the twice saturation density,
the obtained constraints on $E_{\mathrm{sym}}(\rho)$ for $\rho=2\rho_0\sim3\rho_0$ from $R_{\mathrm{nd}}$ are supported by the \emph{ab initio} RBHF and APR calculations~\cite{Akmal1998}.


In summary, the neutron star and the neutron drop have been systematically investigated by \emph{ab initio} calculations as well as the state-of-art DFTs.
Strong correlations are found among the neutron star tidal deformability, the neutron star radius, the root-mean-square radii of neutron drops, and the symmetry energies of nuclear matter at supra-saturation densities.
From these correlations and the tidal deformability extracted from GW170817, the neutron star radii, the neutron drop radii, and the symmetry energy at twice saturation density are respectively constrained as $R_{1.4M_{\odot}}\leqslant 12.94$~km, $R_{\rm nd} \leqslant 2.36$~fm, and $E_{\mathrm{sym}}(2\rho_0) \leqslant 53.2$~MeV.
With the advent of the multi-messenger era, the present work would have an enduring impact on understanding the neutron-rich systems and nuclear EOS with the GW signals.

\begin{acknowledgments}
This work was partly supported by the National Key R\&D Program of China (Contract No. 2017YFE0116700 and No. 2018YFA0404400) and the National Natural Science Foundation of China (NSFC) under Grants No. 11335002, No. 11621131001.
\end{acknowledgments}


%

\end{document}